# Competition ability dependence on uniqueness in some collaboration-competition bipartite networks

Ai-Fen Liu, Xiu-Lian Xu, Chun-Hua Fu, Da-Ren He

*College of Physics Science and Technology, Yangzhou University, Yangzhou, 225002, China*

**Abstract**

Recently, our group quantitatively defined two quantities, "competition ability" and "uniqueness" (Chin. Phys. Lett. 26 (2009) 058901) for a kind of cooperation-competition bipartite networks, where "producers" produce some "products" and "output" them to a "market" to make competition. Factories, universities or restaurants can serve as the examples. In the letter we presented an analytical conclusion that the competition ability was linearly dependent on the uniqueness in the trivial cases, where both the "input quality" and "competition gain" obey normal distributions. The competition between Chinese regional universities was taken as examples. In this article we discuss the abnormal cases where competition gains show the distributions near to power laws. In addition, we extend the study onto all the cooperation-competition bipartite networks and therefore redefine the competition ability. The empirical investigation of the competition ability dependence on the uniqueness in 15 real world collaboration-competition systems is presented, 14 of which belong to the general nontrivial cases. We find that the dependence generally follows the so-called "shifted power law (SPL)", but very near to power laws. The empirically obtained heterogeneity indexes of the distributions of competition ability and uniqueness are also presented. These empirical investigations will be used as a supplementary of a future paper, which will present the comparison and further discussions about the competition ability dependence on the uniqueness in the abnormal collaboration-competition systems and the relationship between the dependence and the competition ability and uniqueness heterogeneity.

## 1. Introduction

The systems where basic elements are simultaneously cooperating and competing are common in nature. If the elements are active, e. g. people or institutions in a society, it is important for them to know the optimum competing strategy. One of our recent letter quantitatively defined two relating quantities, uniqueness and competition ability, and analytically obtained their trivial linear relationship in normal cases.[1] The conclusion proposed that a person or an institution should be as unique as possible to gain the highest competition ability. The cooperation-competition environment was described by a bipartite graph.[1-13] It contains two sets of nodes, one is called "acts" (the "place" for cooperation-competition, such as organizations, activities, or events) and the other is called "actors" (the participants). Edges only exist between different sets of nodes. The degree of an actor node $i$, $h_i$, called "act degree",[8-10] is defined as the act number in which an actor takes part. The degree of an act node $j$, $T_j$, named "act size",[8-10] is defined as the number of the actors, which take part in the act. To describe the cooperation-competition relation between the actors, a projected single-mode (unipartite) network is often used where only the actors exist. In the unipartite network, two actors are connected if they take part in at least one common act.

When quantitatively defined competition ability and uniqueness in Ref. [1], we emphasized that although the idea might be widely applicable, the definitions were only suitable for the following type of systems.[1] The system basic elements ("producers", defined as the actors) "input" some "raw materials", then "manufacture" them and "produce" some "products", and finally "output" ("sell") the products to a "market" (the market for each type of the products are defined as an act where all the producers producing the products are competing). The system basic elements may denote companies, factories, universities (schools) or restaurants and so on. Obviously, this is only one kind of the cooperation-competition bipartite networks. The general cooperation-competition bipartite network actors are not necessarily being producers. The only common characteristic of general cooperation-competition bipartite networks is that some actors compete in some acts for a kind (or several kinds) of resources. For example, the athletes (actors), who join a sport event (an act) in an Olympic Game, are competing for obtaining higher ranking marks. They neither input raw materials nor produce products. This is the first shortage of Ref. [1].

For the kind of the systems discussed in [1], we assumed that the success of a producer in the market depended on both its qualified output quantity, i.e. the number of its selling qualified products (if the number was normalized, it might be regarded as the selling market occupying ratio, or more generally, be regarded as "competition gain", which is defined as an actor obtaining competition resource ratio in an act), and the quality of the raw materials (the input quality, which guaranteed the output quality). Usually it is difficult to quantify the input quality. One can choose a possible indirect description. For example, the price may be chosen for a description of the input quality in a ripe and canonical market. Thus the competition ability was defined in Ref. [1] as $a_i = \sum_{j \in \Gamma}^{h_i}(M_{ij}N_{ij})$ where $\Gamma$ denotes the set of the neighbor acts of actor $i$. In the definition $M_{ij}$ is defined as $M_{ij} = <m_{ij}>/[(\sum_{i \in \Phi}^{T_j} <m_{ij}>)/T_j] = (T_j<m_{ij}>)/(\sum_{i \in \Phi}^{T_j} <m_{ij}>)$ where $\Phi$ denotes the set of the neighbor actors of act $j$, $m_{ij}$ represents the quantified and measurable input quality of actor $i$ in act $j$. The symbol $<>$ represents taking average over all the input raw materials of producer $i$. So $M_{ij}$ denotes the "relative averaged input quality of actor $i$ in act $j$". In the definition, $N_{ij}$ is defined as $N_{ij} = n_{ij}/\sum_{i \in \Phi}^{T_j} n_{ij}$ where $n_{ij}$ represents the qualified output quantity of actor $i$ in act $j$ or more generally, the competition gain of actor $i$ in act $j$. So $N_{ij}$ denotes the "normalized competition gain of actor $i$ in act $j$". The "uniqueness" of actor $i$ is defined as $c_i = \sum_{j \in \Gamma}^{h_i} \frac{1}{T_j}$, which means that the uniqueness becomes increasingly larger if the size of the acts, in which it takes part, decreases and the act number increases. The maximum value of the uniqueness of an actor is $c_{i-\max} = h_{i-\max}$ when $T_j=T_{min}=1$.

In the trivial cases discussed in Ref. [1], both $m_{ij}$ and $n_{ij}$ obey approximate normal distributions, which may be considered as a constant (its average) plus some fluctuations. In these cases we can approximately have $<m_{ij}> \approx m_c$, $n_{ij} \approx n_c$, where $m_c$ and $n_c$ are constants. Thus, the relationship between competition ability and uniqueness is $a_i = \sum_{j \in \Gamma}^{h_i}[(T_j m_c/T_j m_c)(n_c/(T_j n_c))] \approx \sum_{j \in \Gamma}^{h_i}(1/T_j) = c_i$. The relationship indicates that, in normal cases, the competition ability of an actor is almost equal to its uniqueness. Reference [1] only discussed this trivial case. This is the second shortage of Ref. [1].

We will, in the current article and a future paper,[14] extend the investigation presented in Ref,

[1] in two aspects. Firstly, we will discuss all the cooperation-competition bipartite networks without being confined in the special kind, then we have to redefine the competition ability but leave the uniqueness definition unchanged; secondly, we will discuss the nontrivial situations where the distribution of the actor competition ability is abnormal.

As mentioned, the only common characteristic of the general cooperation-competition bipartite networks is that some actors compete in some acts for a kind (or several kinds) of resources. Therefore $M_{ij}$ cannot be defined in the general cases. The competition ability has to be defined as $a'_i = \sum_{j \in \Gamma}^{h_i} N_{ij} = \sum_{j \in \Gamma}^{h_i} (n_{ij} / \sum_{i \in \Phi}^{T_j} n_{ij})$, which means that competition ability of actor $i$ only describes the total competition gain of the actor in the whole network. The uniqueness is still defined as $c'_i = \sum_{j \in \Gamma}^{h_i} \frac{1}{T_j}$. In the nontrivial cases where $N_{ij}$ obeys abnormal distributions, the dependence of the competition ability on the uniqueness should be very different and much more abundant than that in the trivial case. Different general nontrivial systems should show very different dependences. In order to compare the dependences of different systems, we have to normalize both the definitions as $a_i = a'_i / \sum_{i=1}^{N} a'_i$ and $c_i = c'_i / \sum_{i=1}^{N} c'_i$, where $N$ denotes the total number of the actors.

In order to set up a base of further investigations, we will present some important empirical investigations in 15 real world systems in the current long manuscript, which will serve as a supplementary of a future paper.[14] Although several systems belong to the kind studied in [1] (the actors can be regarded as "produces"), the competition ability defined either for the producer-actor type or for the general cases shows abnormal distributions. So, all the 14 systems are nontrivial. Only in one system, which was discussed in [1] (the Zhejiang provincial universities), the competition ability defined for the general cases still shows a normal distribution. From next section on, we will report the investigations in the systems, which are ranked by the decreasing order of the "strength of the competition ability dependence on uniqueness". This will be specified in the future paper.[14]

## 2. Notebook PC selling network at Taobao website

In recent years, on-line shopping by internet becomes more and more popular. Taobao (www.taobao.com) is one of the most famous on-line shopping mall in China. Many shops sell many kinds of commodities through the Taobao website. Thousands of shops sell notebook PCs. The shops collaborate to provide proper notebook PC selling service, and simultaneously compete for more profit. In this bipartite network, the shops are defined as actors, and the selling markets of the notebook PC types are defined as the acts. Usually, the price for the same type of notebook PC can be quite different in different shops. The shops with better reputation can sell out the same type of notebook PCs with higher price. Of course, these shops make more profit. Therefore, the selling price of a type of notebook PCs in a shop can be defined as the competition gain $n_{ij}$ of the actor. Totally 53 notebook PC types and 4711 notebook PC shops in the Taobao on-line shopping mall were collected.[15]

Figures 1 and 2 show that the competition ability and uniqueness distributions can be fitted by so-called "shifted power law (SPL)", which can be expressed as $P(x) \propto (x+\alpha)^{-\gamma}$.[8,10-13] When $\alpha = 0$, it takes a power law form. On the condition that the $x$ is normalized ($0<x_i<1$ and $\sum_{i=1}^{N} x_i = 1$), we can prove that SPL function tends to an exponential function if $\alpha \to 1$.[16] Therefore an SPL interpolates between a power law and an exponential function and the parameter $\alpha$ characterizes the degree of departure from a power law.

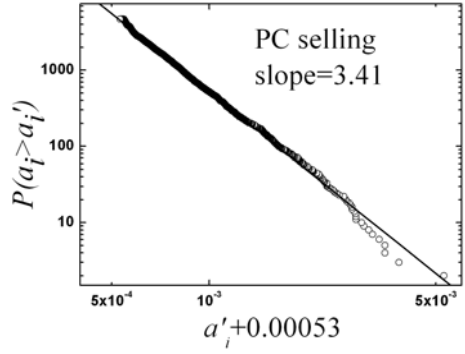

Fig.1: The cumulative distribution of competition ability for Taobao notebook PC selling network

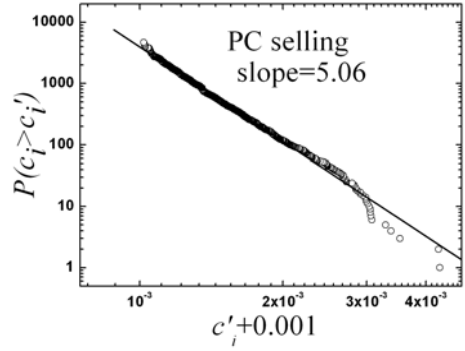

Fig.2: The cumulative distribution of uniqueness for Taobao notebook PC selling network

To describe the heterogeneity of a power law distribution, Hu and Wang defined a "heterogeneity index", $0 \leq H \leq 1$, and deduced an analytic expression of $H$.[17] As mentioned, power law distribution is an extreme case of SPL. Xu et al. extended the Hu-Wang analytic expression from power law to SPL.[13] Consider $N$ values of a quantity, $u$, which are labeled from 1 through $N$ in increasing order of the $u$ value, i.e., $u_1 \leq u_2 \leq \cdots \leq u_N$. An $x$-$y$ plane is defined where $x_i = i/N$ and $y_i = \sum_{j=1}^{i} u_j$. The heterogeneity of $u$ value distribution can be shown by the $y(x)$ line on the $x$-$y$ plane. As shown in Figs. 3 and 4, the heterogeneity index, $H_u$, can be defined as $H_u = S_A/(S_A + S_B) = 1 - 2S_B$,[17,13] where $S_A$ denotes the area between the diagonal line, $y=x$ (which shows that the distribution is absolutely homogeneous), and the $y(x)$ line; and $S_B$ denotes the area beneath the $y(x)$ line, namely the area between the $y(x)$ line, $y=0$ axis, and $x=1$ axis (the two axes indicate that the distribution is completely heterogeneous since $u$ is concentrated on only one value). In figures 3 and 4 the calculation results of the heterogeneity index of both the competition ability distribution and the uniqueness distribution are indicated. The results will be further discussed in [14].

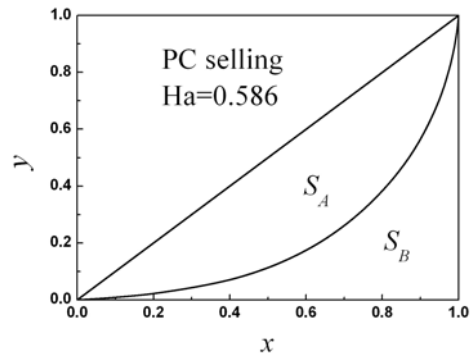

Fig.3: The heterogeneity of the competition ability distribution for Taobao notebook PC selling network

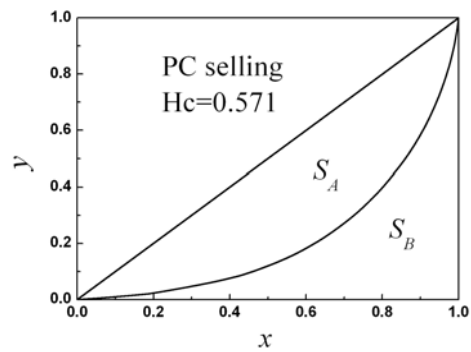

Fig.4: The heterogeneity of the uniqueness distribution for Taobao notebook PC selling network

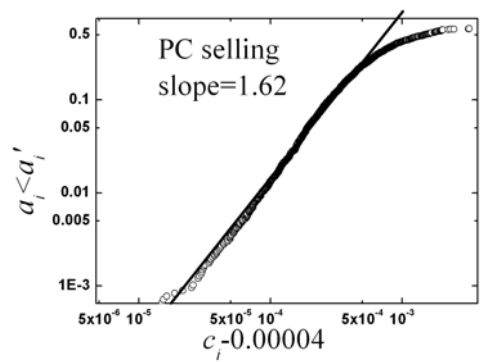

Fig.5: The cumulative node competition ability dependence on its uniqueness for Taobao notebook PC selling network

Figure 5 shows the most important empirical investigation, the cumulative node competition ability dependence on its uniqueness. It also obeys SPL function, which is basically different from a normal distribution. Please note that the SPL function now is expressed as $P(x) \propto (x-\alpha)^{-\gamma}$. It is easy to deduce that the corresponding noncumulative dependence can be approximately expressed as $a_i = 6.6 \times 10^4 (c_i - 4 \times 10^{-4})^{0.62}$ if a quasi-continuous approximation can be accepted and the values of $\alpha$ and the minimum value of $c_i$ are small. Similarly, the function takes a power law form when $\alpha = 0$ and on the condition that the $x$ is normalized ($0 < x_i < 1$ and $\sum_{i=1}^{N} x_i = 1$), we can prove that SPL function tends to an exponential function if $\alpha \to 1$ (see appendix 1). The function is very different from the linear function in the normal case investigated in [1]. In the follows, we will present the dependence functions for other systems. The differences between the functions in all the systems will be further discussed in Ref. [14].

## 3. Beijing restaurant network

Chinese food is famous in the world. Usually, one Chinese restaurant serves a large number of different cooked dishes, and the same cooked dish can be served by many restaurants. We constructed a collaborate-competition bipartite network of the restaurants in Beijing, in which the restaurants are defined as actors and the selling markets of the cooked dishes are defined as acts. In addition to collaborating to provide the food services, the restaurants serving the same type of dishes also compete to attract more customers and therefore earn more profits. Actually, this system belongs to the "producer actor type" as discussed in [1], however, the competition ability shows abnormal distributions with either the definition in [1] or the definition in the current manuscript for the general cases, therefore we now regard the system as a general one. In the definition stated in Sec. 1, $n_{ij}$ represents the "attention degree", which is quantified by the customer marks given in the "Dianping website" (www.dianping.com/beijing), for dish $j$ in restaurant $i$. Until 2006, we collected 688 cooked dishes in 3337 restaurants.[15] Figures 6-10 show the empirical investigations.

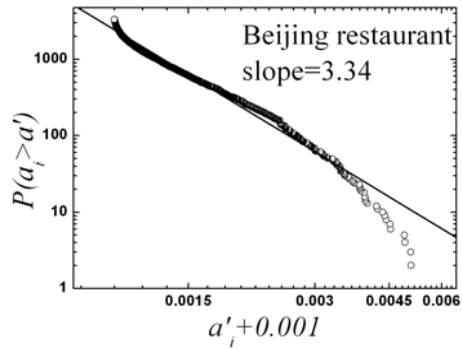

Fig.6: The cumulative distribution of competition ability for Beijing restaurant network

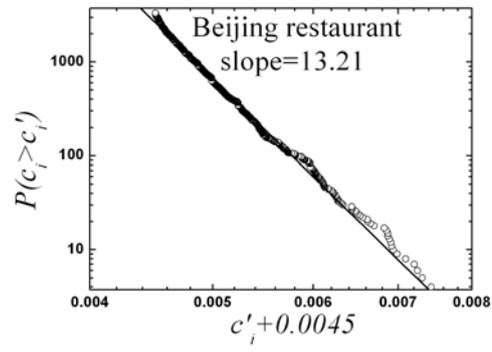

Fig.7: The cumulative distribution of uniqueness for Beijing restaurant network

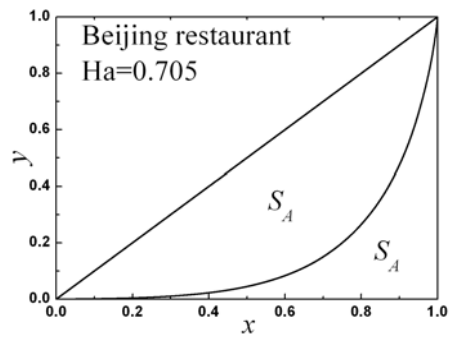

Fig.8: The heterogeneity of the competition ability distribution for Beijing restaurant network

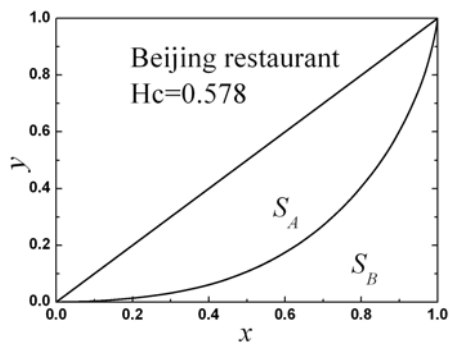

Fig.8: The heterogeneity of the uniqueness distribution for Beijing restaurant network

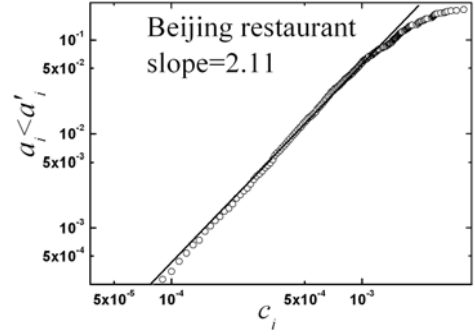

Fig.10: The node competition ability dependence on its uniqueness for Beijing restaurant network

The corresponding noncumulative dependence can be approximately expressed as $a_i = 2.5 \times 10^5 \times c_i^{1.11}$.

## 4. University matriculation network

In Chinese universities, colleges or departments are divided into undergraduate specializations. The undergraduates of different specializations are taught by different teaching schemes. In order to enter a specialization, a middle school student has to pass the national matriculation with the total marks higher than a lowest value. The university recruitment process is divided into several (roughly speaking, 6) batches in each province or region. Different batches have different lowest values of the matriculation mark. Different specializations may have the recruitment right in different batches depending on their academic levels. Also, each specialization may have the recruitment right in different batches in different geographical regions. A specialization with a higher batch recruitment right can recruit higher mark students. Therefore, universities are competing in the recruitment process in many "batches", which depend on specialization and geographical region, to recruit more and better students. From a different view point, all the universities are also cooperating in the total recruitment process to successfully complete the well-organized job. We construct the university matriculation network by defining the batches as the acts, universities as the actors, and the lowest matriculation marks as the competition gain, $n_{ij}$, respectively. There are 51 batches and 2277 universities. The data were downloaded from the web stations www.hneeb.cn; www.jszs.net; www.lnzsks.com; and www.nm.zsks.cn.[15] Figures 11-15 show the empirical investigations.

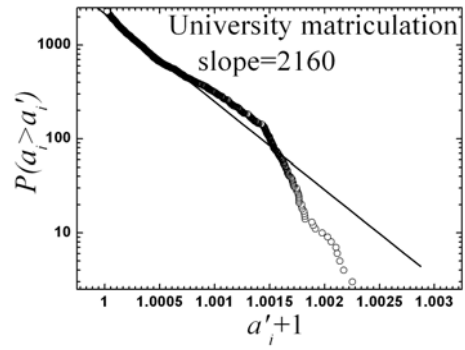

Fig.11: The cumulative distribution of competition ability for university matriculation network

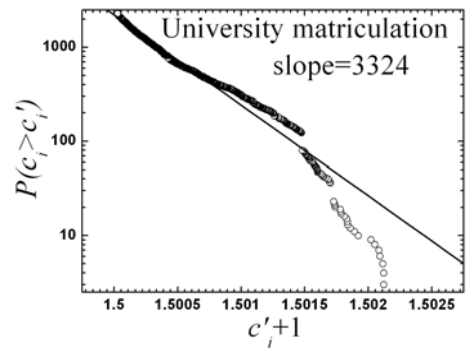

Fig.12: The cumulative distribution of uniqueness for university matriculation network

Figures 11 and 12 actually show that the competition ability and the uniqueness distributions obey so-called "double SPL functions", which means that in each figure the data have to be fitted by two SPL functions. One fits the higher half of the data, the other fits another half. In the figures only the slope of the first SPL fitting is indicated for a rough comparison of the competition ability dependence on uniqueness in the 15 systems, which will be reported in Ref. [14].

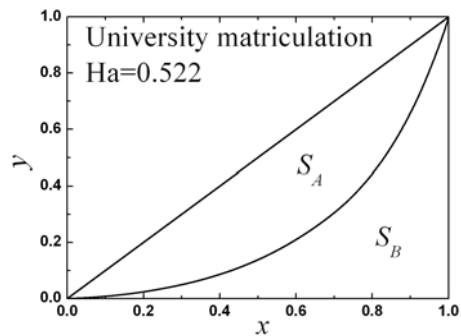

Fig.13: The heterogeneity of the competition ability distribution for university matriculation network

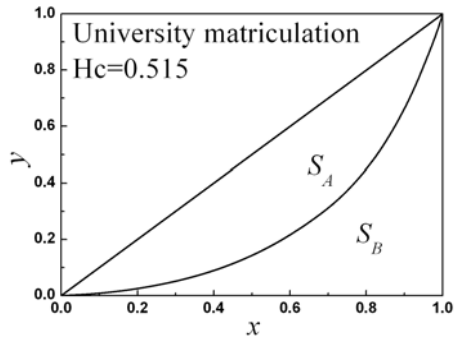

Fig.14: The heterogeneity of the uniqueness distribution for university matriculation network

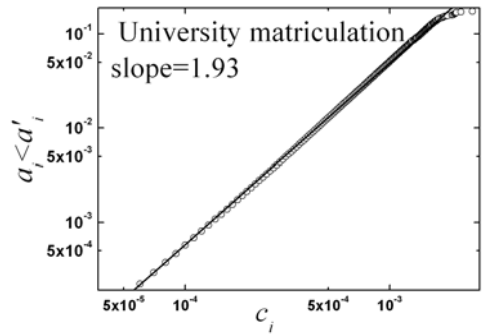

Fig.15: The node competition ability dependence on its uniqueness for university matriculation network

The corresponding noncumulative dependence can be approximately expressed as $a_i = 5.8 \times 10^4 \times c_i^{0.93}$.

## 5. University independent recruitment network

In China, all the universities can recruit new students from middle schools only via a nationwide uniform activity. All the middle school students have to pass the national matriculation. The delegacies of different academic level universities then get together to select middle school students according to the matriculation marks of the students and the university batch rights just as introduced in last section. However, in recent years, some best universities are awarded the right to recruit middle school students before the national matriculation. They can perform their own examinations or interviews and then recruit new students by their own decision. We call this as "independent recruitment". The middle schools can achieve better reputations if more of their students have been recruited by these top level universities. Therefore, the middle schools collaborate to accomplish the independent recruitment activity and simultaneously compete for sending more students to the universities. We define the middle schools as the actors, the independent recruitment universities as the acts, and the number of the recruited students as the competition gain. In 2006, there were 53 universities, which had the independent recruitment right. However, only the data of 52 were available. 1546 middle schools and their recruited student numbers were included in the data (from www.chsi.com.cn).[15] Figures 16-20 present the empirical investigations.

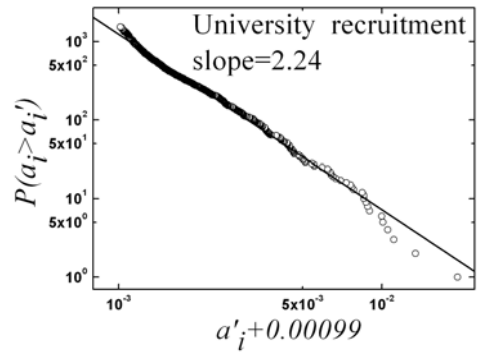

Fig.16: The cumulative distribution of competition ability for university independent recruitment network

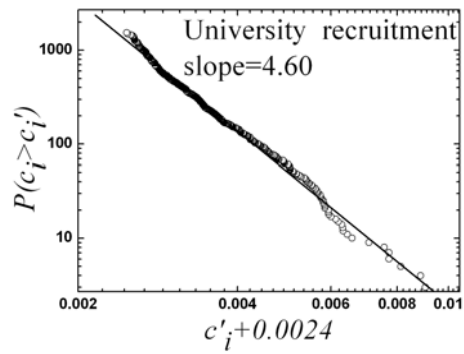

Fig.17: The cumulative distribution of uniqueness for university independent recruitment network

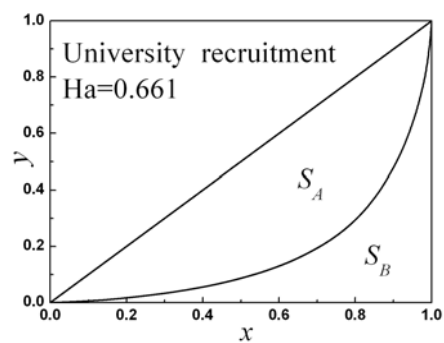

Fig.18: The heterogeneity of the competition ability distribution for university independent recruitment network

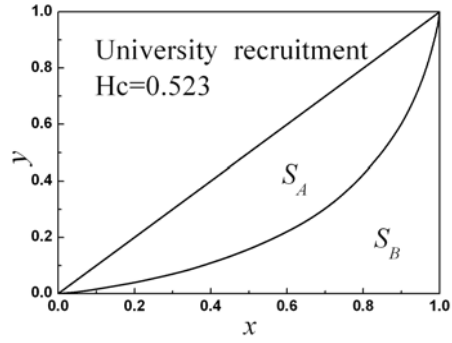

Fig.19: The heterogeneity of the uniqueness distribution for university independent recruitment network

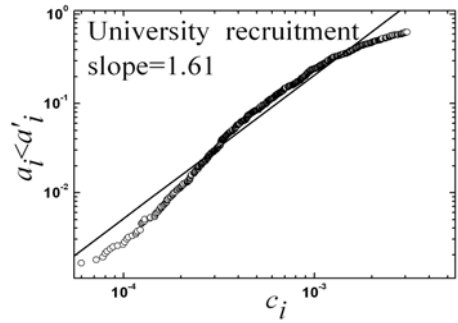

Fig.20: The node competition ability dependence on its uniqueness for university independent recruitment network

The corresponding noncumulative dependence can be approximately expressed as $a_i = 2.3 \times 10^4 \times c_i^{0.61}$.

## 6. Information technique product selling network

The market of information technique (IT) products, including mobile phone, computer, digital camera, and so on, is another example which shows typical collaboration-competition characteristics. If different manufacturers produce the same type of IT products, they compete in the selling market. Of course, these manufacturers also collaborate to supply enough IT products, and to maintain the market order. We construct a collaboration-competition bipartite network in which the manufactures are defined as actors and the selling market of the IT products are defined as acts. On the website www.pcpop.com, there are detailed introductions to each IT product produced by a specific manufacturer. This web site also gives the "attention rank" of the manufacturers for each IT product according to the total browsing time by the customers. Actually, this system is a typical sample of the "producer actor type" as discussed in [1], however, the competition ability again shows abnormal distributions with either the definition in [1] or in the current manuscript for the general cases, therefore we also regard the system as a general one. In the competition ability definition $n_{ij}$ represents the "attention rank", which is relevant to the competition abilities of the manufactures, and can be used to quantify the competition achievement, i.e., the profit. We collected 265 manufacturers and 2121 IT products from the website.[15] Figs. 21-25 show the empirical investigations.

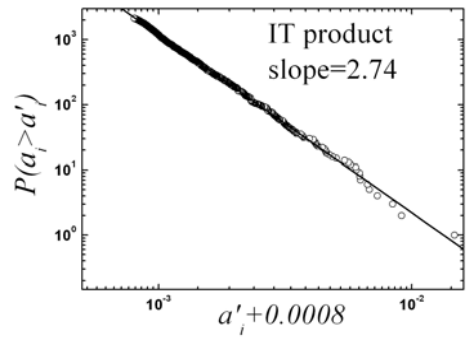

Fig.21: The cumulative distribution of competition ability for IT product selling network

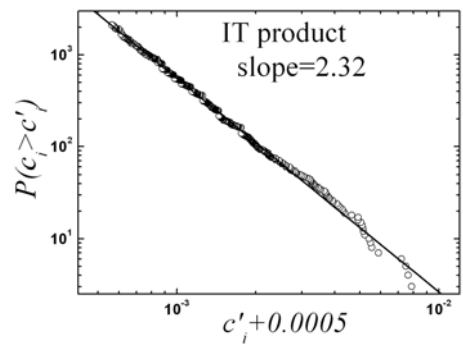

Fig.22: The cumulative distribution of uniqueness for IT product selling network

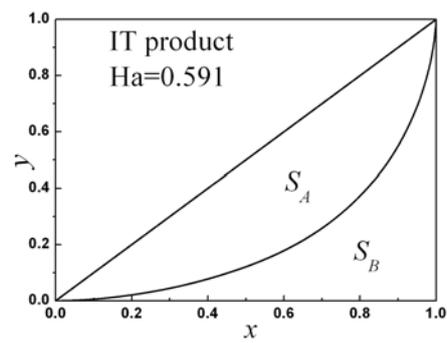

Fig.23: The heterogeneity of the competition ability distribution for IT product selling network

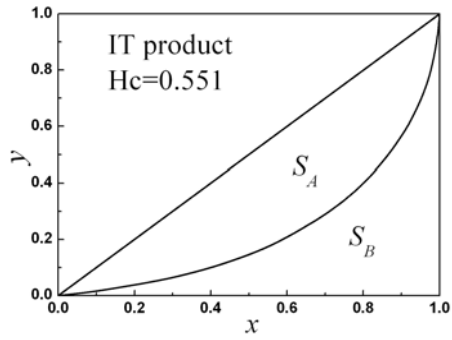

Fig.24: The heterogeneity of the uniqueness distribution for IT product selling network

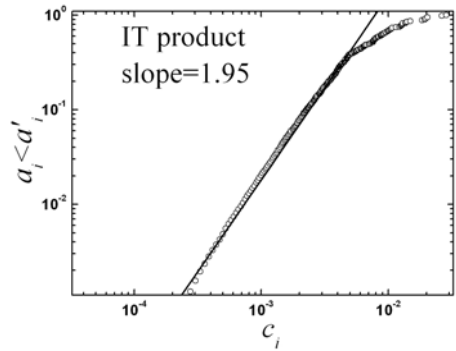

Fig.25: The node competition ability dependence on its uniqueness for IT product selling network

The corresponding noncumulative dependence can be approximately expressed as $a_i = 2.3 \times 10^4 \times c_i^{0.95}$.

## 7. 2004 Athens Olympic network

In an Olympic game some athletes join a sport event to successfully conduct the pageant and also to obtain more sport scores. We construct the network of the 2004 Athens Olympic Game by defining the athletes as the actor nodes and the sport events (only the individual sport events, e.g., high jump, weight lifting, are considered) as acts. The data were downloaded from www.sina.com.cn (2004), which includes 133 individual sport events and 4500 athletes, as well as their sport scores in each sport event.[15] The sport scores are defined as the competition gain. Figs. 26-30 show the empirical investigations.

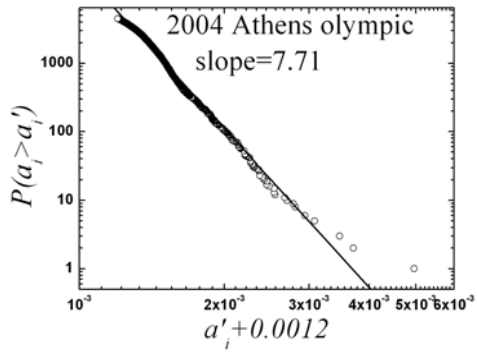

Fig.26: The cumulative distribution of competition ability for 2004 Athens Olympic network

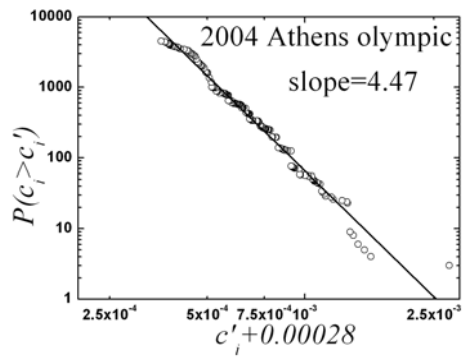

Fig.27: The cumulative distribution of uniqueness for 2004 Athens Olympic network

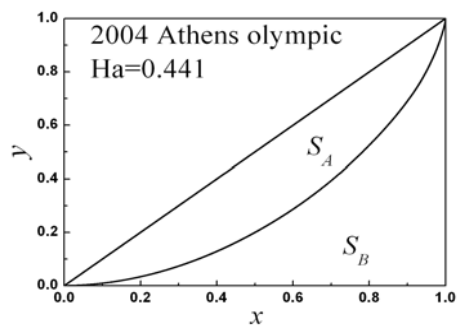

Fig.28: The heterogeneity of the competition ability distribution for 2004 Athens Olympic network

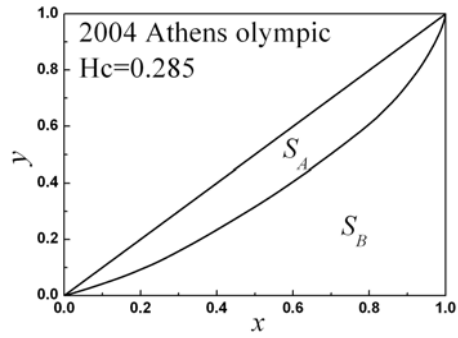

Fig.29: The heterogeneity of the uniqueness distribution for 2004 Athens Olympic network

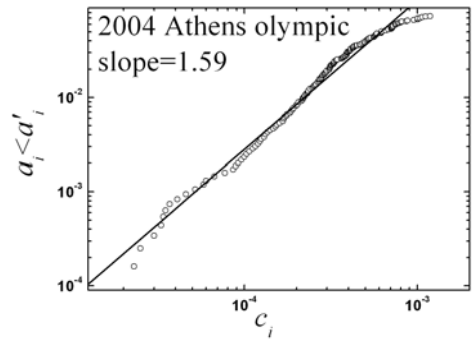

Fig.30: The node competition ability dependence on its uniqueness for 2004 Athens Olympic network

The corresponding noncumulative dependence can be approximately expressed as $a_i = 9.8 \times 10^3 \times c_i^{0.59}$.

## 8. China mainland movie network

The data were downloaded from the webset (http://www.movdb.com). Based on the data, we constructed a China mainland movie network in which the movies are defined as actors and the leading movie actors as acts. The movies acted by the same leading actors often belong to a common movie type, which has a certain type of audience. In this sense, these movies compete to attract more audience in addition to collaborating to provide abundant entertainments. More downloading numbers of a movie indicates more audience and therefore more ticket office income. We therefore define the downloading numbers as the competition gain. We collected the data for 3084 movies involving 920 leading actors before April 29, 2007.[15,18] Figs.31-35 show the empirical investigations.

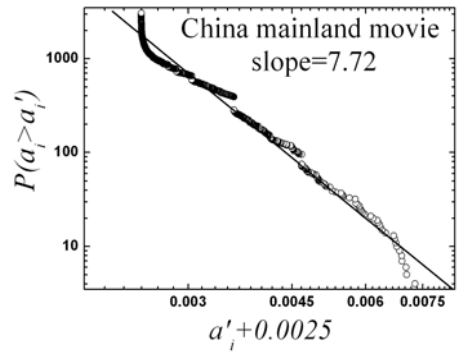

Fig.31: The cumulative distribution of competition ability for China mainland movie network

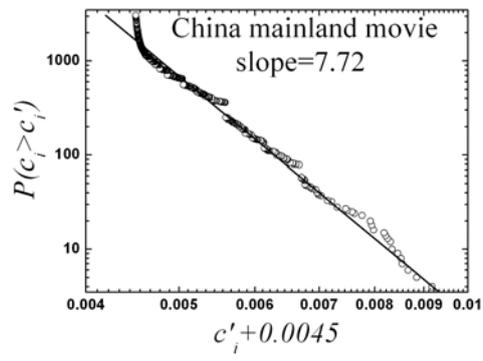

Fig.32: The cumulative distribution of uniqueness for China mainland movie network

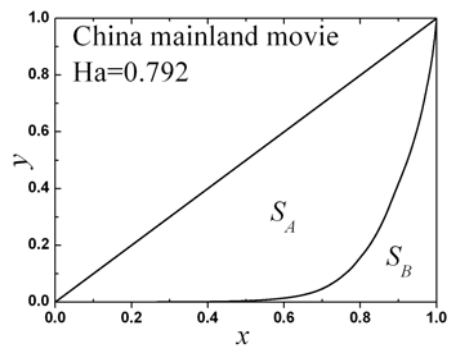

Fig.33: The heterogeneity of the competition ability distribution for China mainland movie network

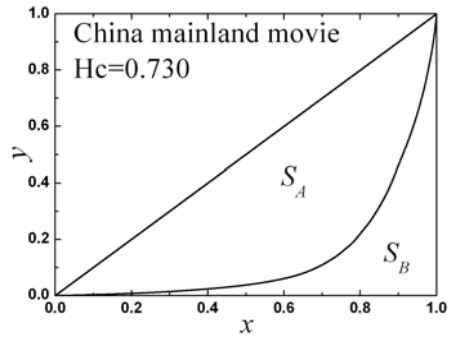

Fig.34: The heterogeneity of the uniqueness distribution for China mainland movie network

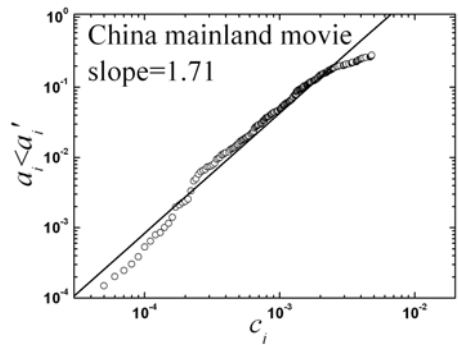

Fig.35: The node competition ability dependence on its uniqueness for China mainland movie network

The corresponding noncumulative dependence can be approximately expressed as $a_i = 1.0 \times 10^4 \times c_i^{0.71}$.

## 9. Undergraduate course selection network of Yangzhou University

Yangzhou University (YZU) is a rather new university that was founded in 1992 by the union of 7 smaller colleges. Although its history is short, it develops quite quickly in recent years. In 2006, in its 27 colleges, YZU offers 98 undergraduate programs covering 11 disciplines for 33900 undergraduates. In addition to the courses typically presented in every college, YZU also offers 121 general support courses between 2002 and 2006 (our data does not include the courses after 2006), which cover all the natural and social scientific disciplines. Every undergraduate has to take at least 4 selective courses among 121 ones. These 121 general support courses can be regarded as the parts of 78 scientific subjects such as physics, mathematic, art and so on. One course may belong to more than one scientific subject. Based on the course selection data of 65,536 undergraduates we built an undergraduate course selection network of YZU.[15] We define the general support courses as the actors. Some certain actors are related to each other by being selected by the students from the same scientific subject (these scientific subjects are defined as the acts of the bipartite network). The competition gain is defined as the number of the undergraduates who take this selective general support course. One may say that the general support courses cooperate to form the general education system in YZU, and also compete for attracting more students. Figures 36-40 show the empirical investigations.

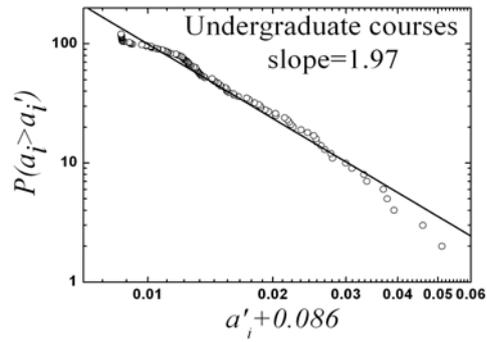

Fig.36: The cumulative distribution of competition ability for undergraduate course selection network of YZU

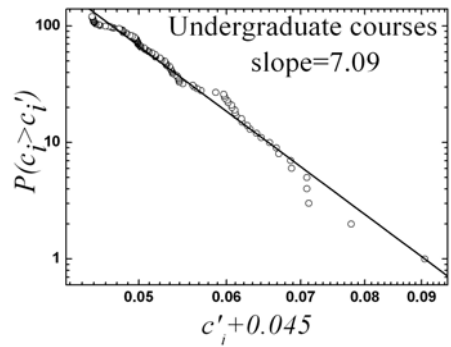

Fig.37: The cumulative distribution of uniqueness for undergraduate course selection network of YZU

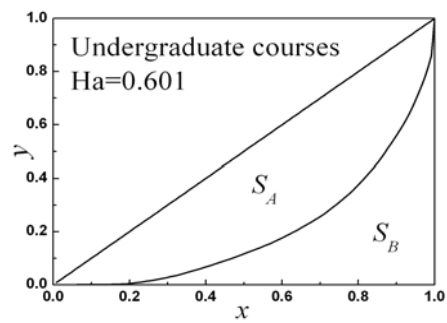

Fig.38: The heterogeneity of the competition ability distribution for undergraduate course selection network of YZU

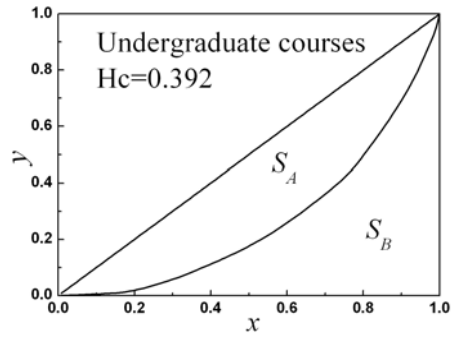

Fig.39: The heterogeneity of the uniqueness distribution for undergraduate course selection network of YZU

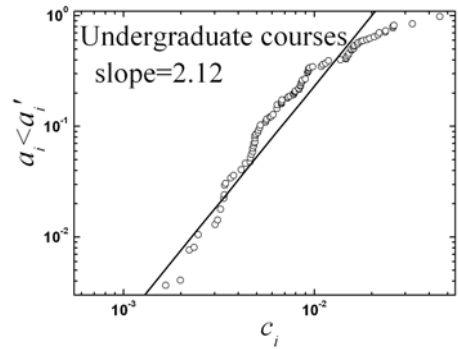

Fig.40: The node competition ability dependence on its uniqueness for undergraduate course selection network of YZU

The corresponding noncumulative dependence can be approximately expressed as $a_i = 8.6 \times 10^3 \times c_i^{1.12}$.

## 10. Book borrowing network of YZU library

We also constructed a book borrowing network of YZU library. In library, some books are very popular and have been borrowed frequently by the readers. In comparison, some books have never been borrowed. Such kind of book borrowing information is very important for evaluating the importance of a certain book and is useful for assessing the knowledge configuration of the students. In YZU library, there are 15204 physics books with some books having the same name (for example, for same important subjects, there are usually more than 30 books having the same book name). We collected the borrowing record data for 3207 books with nonzero borrowing record before 2006. These books are further classified into 227 different scientific subjects (such as electromagnetism, quantum mechanics, and so on). One subject includes many different books, and one book may belong to more than one scientific subject. In constructing the bipartite network, we define the books as actors and the scientific subjects as acts. The actors (books) cooperate in a certain act (scientific subject) to provide the necessary knowledge, and also compete for being borrowed more frequently. Therefore, we define the competition gain $n_{ij}$ as the number of borrowing records. Totally 88981 records are used in constructing the network.[15] Figs. 41-45 show the empirical investigations.

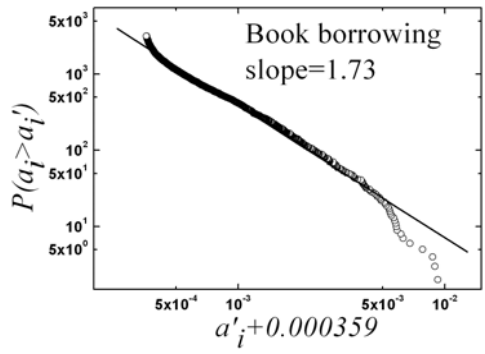

Fig.41: The cumulative distribution of competition ability for book borrowing network of YZU library

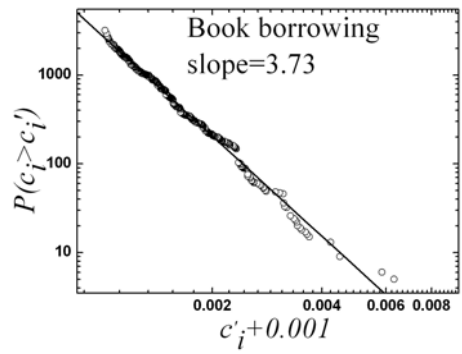

Fig.42: The cumulative distribution of uniqueness for book borrowing network of YZU library

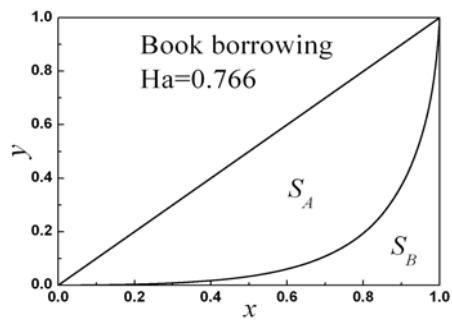

Fig.43: The heterogeneity of the competition ability distribution for book borrowing network of YZU library

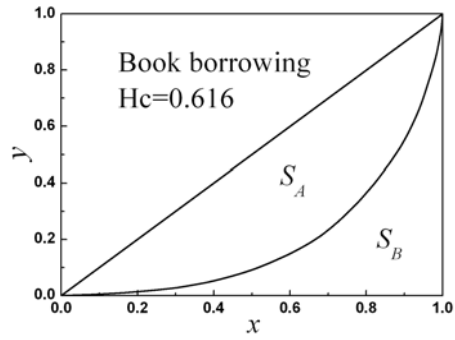

Fig.44: The heterogeneity of the uniqueness distribution for book borrowing network of YZU library

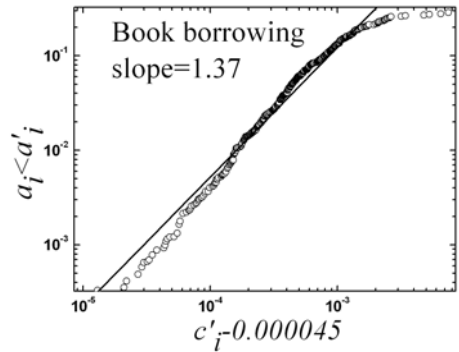

Fig.45: The node competition ability dependence on its uniqueness for book borrowing network of YZU library

The corresponding noncumulative dependence can be approximately expressed as $a_i = 2.2 \times 10^3 \times (c_i - 4.5 \times 10^{-5})^{0.37}$.

## 11. Supermarket network

Supermarkets selling the same commodities collaborate to provide better services to the customers and compete to attract more buyers for more profits. In a web site, www.dianping.com, the buyers give marks for each commodity in every supermarket according to the supermarket environment, service and each commodity quality and price. For a supermarket, higher marks mean more buyers and profits. We define the supermarkets as the actors, the commodities as the acts, and the buyer's marks as the competition gain $n_{ij}$. Totally 1046 kinds of commodities and 570 supermarkets are included in our data.[15] Figs.46-50 show the empirical investigations.

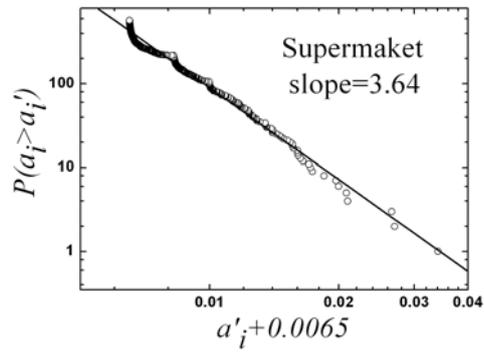

Fig.46: The cumulative distribution of competition ability for supermarket network

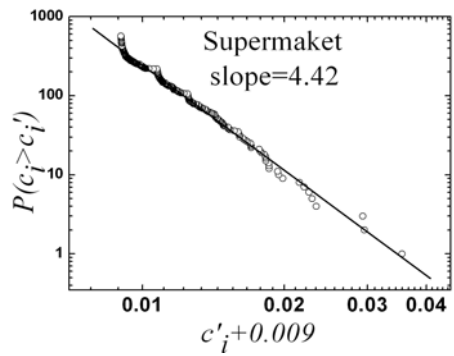

Fig.47: The cumulative distribution of uniqueness for supermarket network

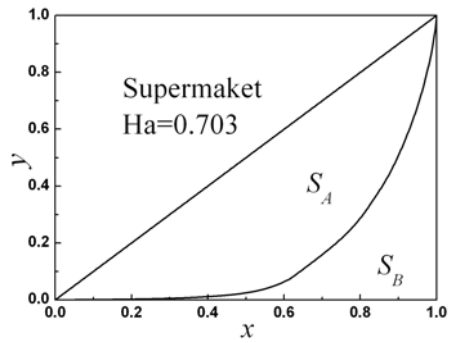

Fig.48: The heterogeneity of the competition ability distribution for supermarket network

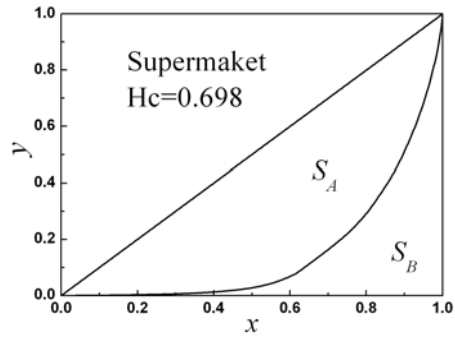

Fig.49: The heterogeneity of the uniqueness distribution for supermarket network

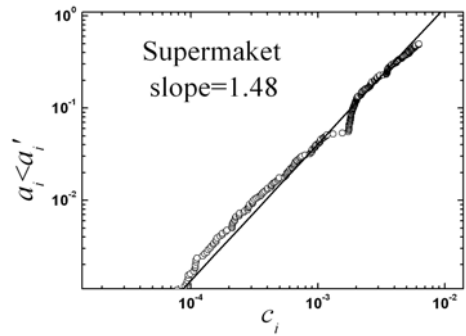

Fig.50: The node competition ability dependence on its uniqueness for supermarket network

The corresponding noncumulative dependence can be approximately expressed as $a_i = 1.7 \times 10^3 \times c_i^{0.48}$.

## 12. Training institution network

In recent 20 years Chinese employment market becomes very active. As the result, there have appeared many training institutions which offer many different training courses. The institutions providing the same training courses cooperate to form the proper market, and also compete for more tuition fee income. In the network, the training institutions are defined as the actors and the training courses are defined as the acts. The tuition fee for a certain training course is defined as the competition gain $n_{ij}$. We collected the data of 398 training institutions in China, which are authorized by the national ministry of education and/or the national labor bureau. Until the data collection, 2006, totally 2673 training courses are included in the data (from www.ot51.com, www.00100.cc, www.people.com.cn et al.). [15] Figures 51-55 show the empirical investigations.

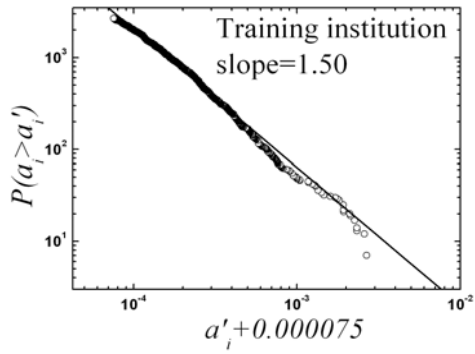

Fig.51: The cumulative distribution of competition ability for training institution network

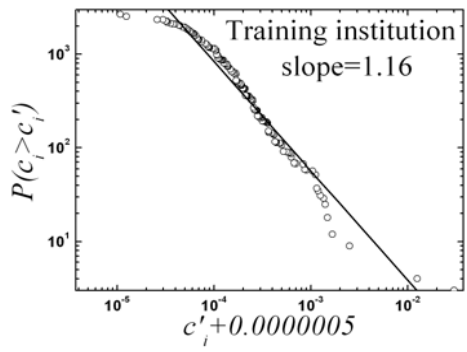

Fig.52: The cumulative distribution of uniqueness for training institution network

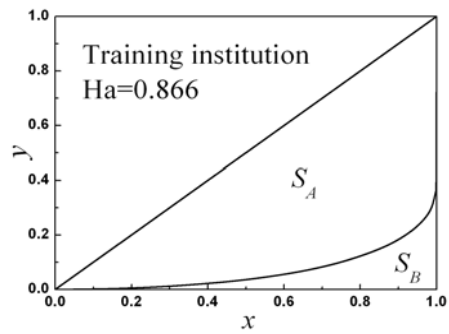

Fig.53: The heterogeneity of the competition ability distribution for training institution network

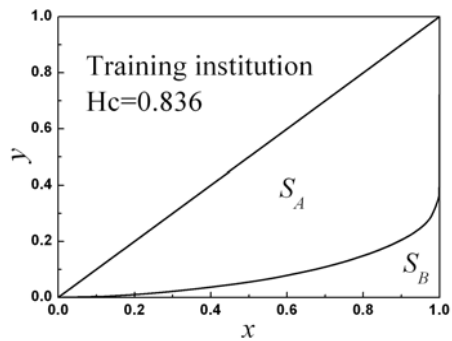

Fig.54: The heterogeneity of the uniqueness distribution for training institution network

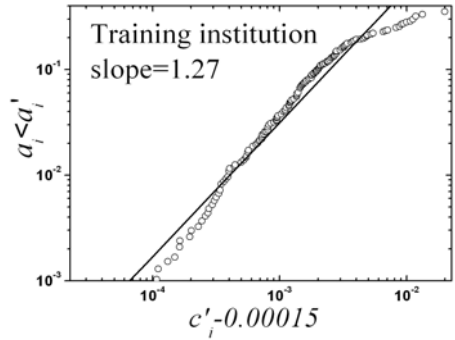

Fig.55: The node competition ability dependence on its uniqueness for training institution network

The corresponding noncumulative dependence can be approximately expressed as $a_i = 2.5 \times 10^2 \times (c_i - 1.5 \times 10^{-4})^{0.27}$.

## 13. Human acupuncture point network

Acupuncture is one of the most important traditional Chinese therapy means. Because of its practical efficacy in curing some diseases, acupuncture becomes more and more popular over the world. There are 187 key acupuncture points in human body. To treat a certain disease, several specific acupuncture points need to be acupunctured simultaneously. In this sense, the acupuncture points collaborate in curing a certain disease. Therefore, we can define the diseases treated by acupuncture as acts, and the acupuncture points as actors. However, for a certain disease, not all of the relevant acupuncture points play the equal role. Some acupuncture points are more crucial. Usually, these relatively more important acupuncture points, in treating a certain disease, need more acupuncture times. Therefore, we define the acupuncture time as the competition gain $n_{ij}$. Totally 108 different kinds of diseases and 187 acupuncture points are involved in our data (from www.acutimes.com ).[15] Figures 56-60 show the empirical investigations.

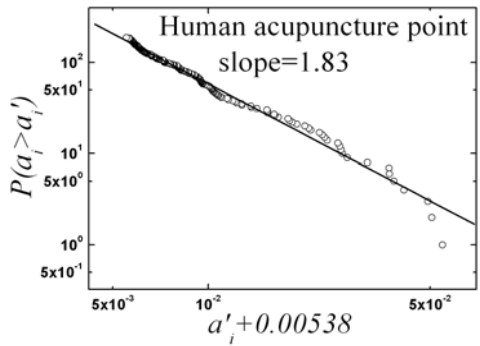

Fig.56: The cumulative distribution of competition ability for human acupuncture point network

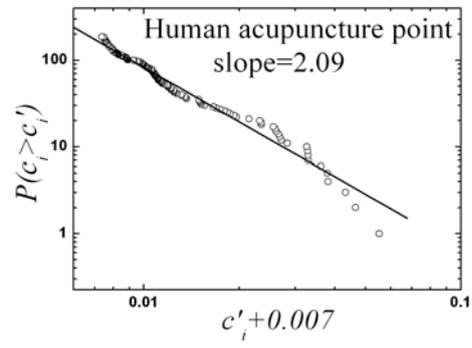

Fig.57: The cumulative distribution of uniqueness for human acupuncture point network

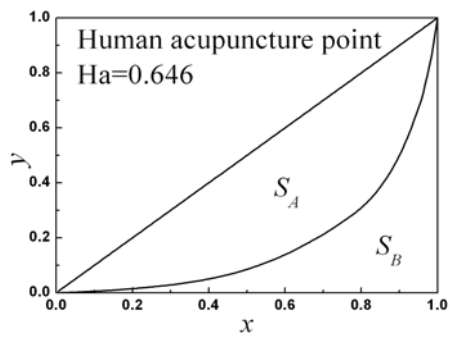

Fig.58: The heterogeneity of the competition ability distribution for human acupuncture point network

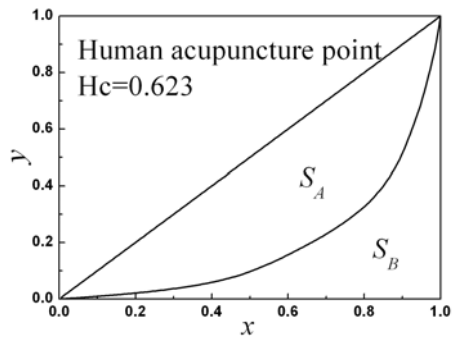

Fig.59: The heterogeneity of the uniqueness distribution for human acupuncture point network

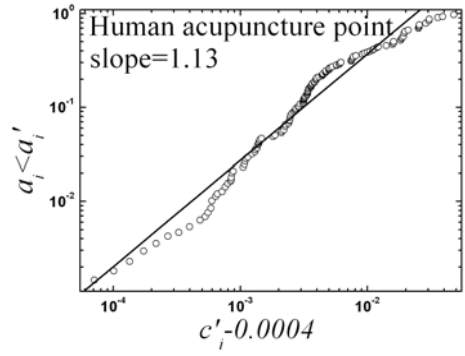

Fig.60: The node competition ability dependence on its uniqueness for human acupuncture point network

The corresponding noncumulative dependence can be approximately expressed as $a_i = 78.2 \times (c_i - 4.0 \times 10^{-4})^{0.13}$.

## 14. Mixed drink network

    The mixed drinks (such as cocktails) usually contain a large number of ingredients according to consumer's taste, and many mixed drinks may share the same ingredients. We construct the mixed drink network by defining the component ingredients as actors and the mixed drinks as acts. The ingredients collaborate to form mixed drinks with different tastes. Simultaneously, the ingredients contained in a common mixed drink can be regarded as being competing since the ingredients have different relative importance. As the first step investigation, we very simply suppose that a certain ingredient in higher proportion is relatively more important. Therefore, we define the relative proportion of each ingredient in a certain mixed drink as the competition gain $n_{ij}$. Until 2006, we collected 7804 mixed drinks and 1501 ingredients. The proportions of the ingredients in each mixed drink are also obtained (www.drinknation.com).[15] Figures 61-65 show the empirical investigations.

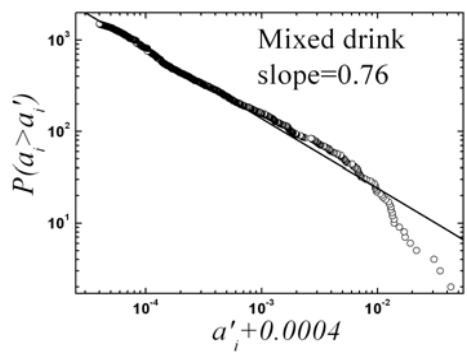

Fig.61: The cumulative distribution of competition ability for mixed drink network

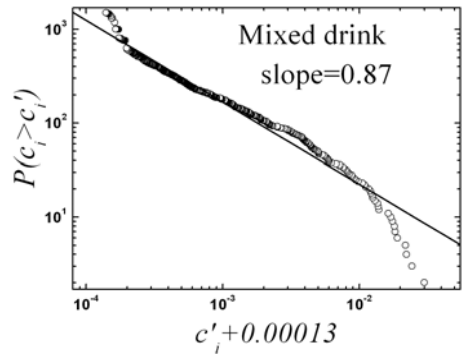

Fig.62: The cumulative distribution of uniqueness for mixed drink network

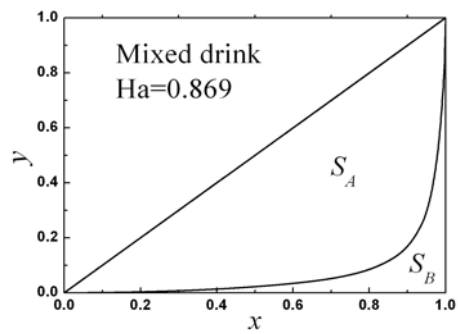

Fig.63: The heterogeneity of the competition ability distribution for mixed drink network

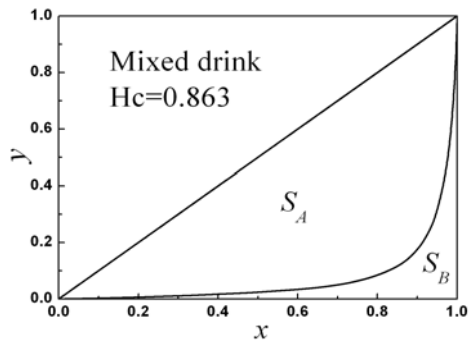

Fig.64: The heterogeneity of the uniqueness distribution for mixed drink network

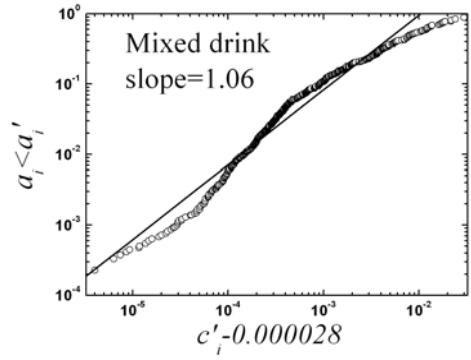

Fig.65: The node competition ability dependence on its uniqueness for mixed drink network

The corresponding noncumulative dependence can be approximately expressed as $a_i = 56.9 \times (c_i - 2.8 \times 10^{-5})^{0.06}$.

## 15. Zhejiang provincial university network

In Ref. [1], we presented empirical investigations on the competition between so-called "regional universities" in China as proofs of the analytically obtained linear competition ability dependence on uniqueness for the trivial cases of the "producer actor type of systems". Universities in four provinces: Jiangsu, Zhejiang (on behalf of the circumlittoral developed area in China) and Shaanxi, Ningxia (on behalf of the central and western developing area in China) were studied.

In the last 10 years, Chinese higher-education has experienced a very unusual scale expansion. In particular, the regional universities, which got outlay from and belonged to provincial governments, have recruited many more students than 10 years ago. The regional universities located in the same province have to compete with each other for the resources of better high school students and larger portions in the employment market. Since they are comparable in quality, the competition is severe. In these systems, the regional universities can be defined as actors, and the undergraduate specializations can be defined as acts. In Chinese universities, colleges or departments are divided into undergraduate specializations. The undergraduates of different specializations are taught by different teaching schemes. They usually search for different kinds of jobs when they graduate. In order to enter a specialization, a high school student has to pass the national matriculation with the total marks higher than a value defined by the specialization.

In the current manuscript we discuss general competing systems, which are not necessarily producer actor type. In order to compare the regional university competition systems with the general ones introduced in the previous sections, we present in this section an empirical investigation on Zhejiang universities but with the general definitions of the competition ability and the uniqueness as mentioned in Sec. 1. We define the student number enrolled in a university specialization as the competition gain $n_{ij}$, assuming that this number is proportional to the number of the graduated students who find jobs (which can be regarded as the competition gain, but it is difficult to get the data). Zhejiang has 24 regional universities and 158 specializations.[19] The student number enrolled in each university specialization can also be found in [19]. Figures 66-70 show the empirical investigations.

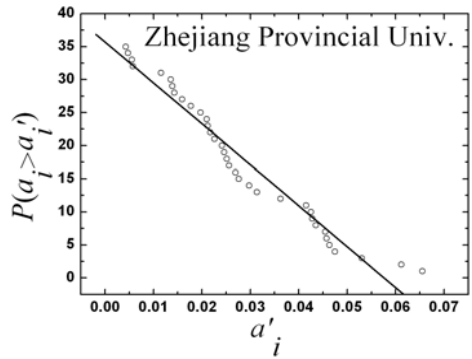

Fig.66: The cumulative distribution of competition ability for Zhejiang provincial university network

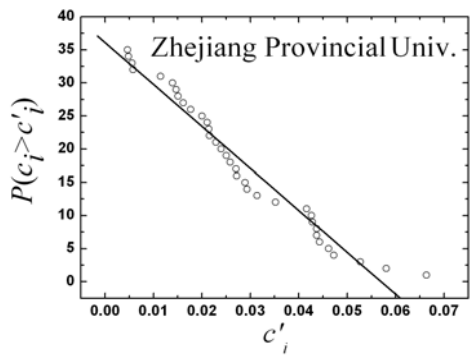

Fig.67: The cumulative distribution of uniqueness for Zhejiang provincial university network

As can be seen in figures 66 and 67, the cumulative distributions of the competition ability and the uniqueness show linear functions, which indicated the corresponding even noncumulative distributions, or we can say that both the competition ability and the uniqueness can be regarded as constants. This is in agreement with the conclusion reported in Ref. [1] with the old definition.

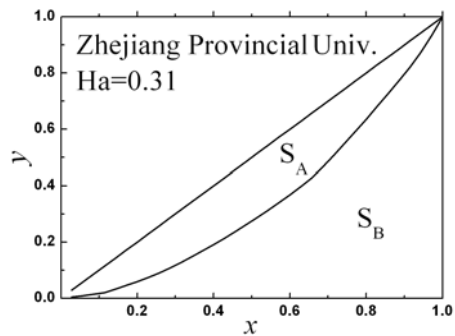

Fig.68: The heterogeneity of the competition ability distribution for Zhejiang provincial university network

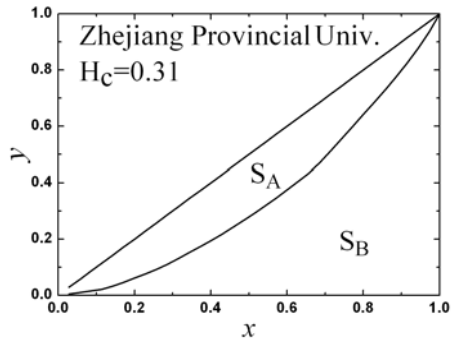

Fig.69: The heterogeneity of the uniqueness distribution for Zhejiang provincial university network

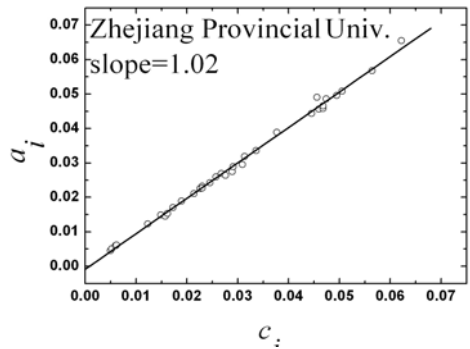

Fig.70: The node competition ability dependence on its uniqueness for Zhejiang provincial university network.

Figure 70 clearly shows that the node competition ability dependence on its uniqueness for Zhejiang provincial university network obeys a linear function $a_i=c_i$ with the general definitions.

## 16. World language distribution network

Based on the data from Ethnologue website (http://www.ethnologue.com, the 15[th] edition, published in 2005) which shows the populations speaking each language in a country (or a relatively independent region), we constructed a world language distribution network. The languages are defined as actors, and the countries are defined as acts. In the bipartite graph, if a certain language is used in a country, this language and the country are connected by an edge. In the projected unipartite graph, the edge between two languages means that the languages are used simultaneously at least in one common country. In a very long time consideration, the languages can be considered as being competing to be used by more people. As the result, some languages have died out, but some other languages have been spoken by more and more people and spread to more and more geographical regions. The languages also collaborate in the common regions to accomplish the communications between the people. Therefore, we define the populations speaking a certain language in the country as the competition gain $n_{ij}$. We collected 6142 kinds of languages used in 236 countries and regions. The total number of the language speakers is 5.2385 $\times 10^9$. Since the Ethnologue data are taken from multiple sources, the sum of the languages' population may not exactly equal to the total population in the world. Figures 71-75 show the empirical investigations.

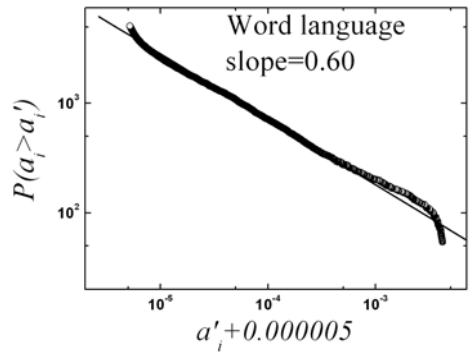

Fig.71: The cumulative distribution of competition ability for word language distribution network

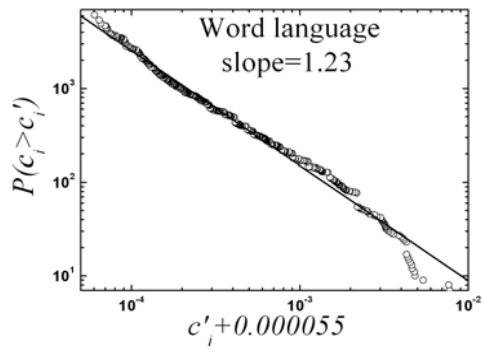

Fig.72: The cumulative distribution of uniqueness for word language distribution network

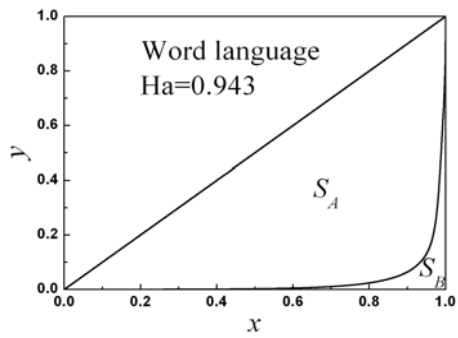

Fig.73: The heterogeneity of the competition ability distribution for word language distribution network

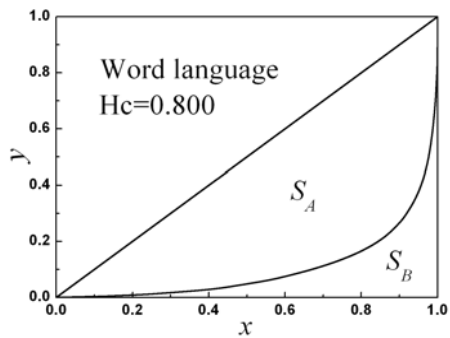

Fig.74: The heterogeneity of the uniqueness distribution for word language distribution network

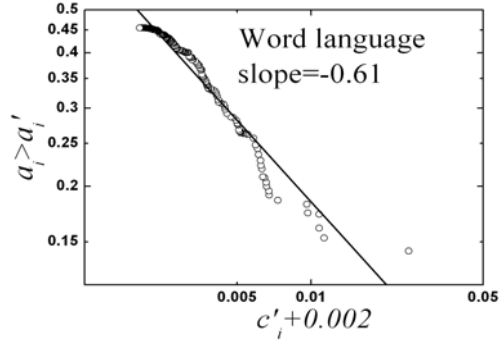

Fig.75: The node competition ability dependence on its uniqueness for word language distribution network

Figure 75 shows a very unique competition ability dependence on uniqueness, which shows a negative slope relationship. This means that the competition ability becomes increasingly smaller when the uniqueness increases. We shall discuss this further in Ref. [14]. The corresponding noncumulative dependence can be approximately expressed as $a_i = 7.0 \times 10^{-3}(c_i - 2.0 \times 10^{-3})^{-1.61}$.

## 17. Summary

In this long manuscript we present the empirical investigations of the distributions of competition ability and uniqueness and on the heterogeneities of the distributions. The empirical investigations of the competition ability dependence on the uniqueness are also reported. 15 real world cooperation-competition networks are involved.

Almost in all the systems the data of the distributions of competition ability and uniqueness can be well fitted by SPL functions. The only exception is shown in university matriculation network where the data should be fitted by two SPL functions but only the first one is mentioned. This does not matter since the distributions will be used only to show two things in the future paper.[14] The first one is that the distributions are abnormal, therefore the corresponding investigation on the competition ability dependence on the uniqueness is beyond the trivial cases; the second role of the distributions is to provide the heterogeneity information of them so that the relationship between the strength of the competition ability dependence on the uniqueness and the distribution heterogeneity can be discussed. The rough description of the distributions in university matriculation network does not influence either of the functions.

In four systems, the university independent recruitment network, YZU undergraduate course selection network, training institution network and the word language distribution network, the data are fairly well fitted by SPL functions with a minus $\alpha$ parameter. In appendix 1 we explain that this function is also an interpolation between power law and exponential decay. In all the other systems the data are fitted very well. Zhejiang university network shows a special case where both the SPL parameters show special values, $\alpha=0$ and $\gamma=1$. The word language distribution network is another exception, which shows a negative $\gamma$ value. All the information will be very important in our future paper [14] for a discussion of the different strengths of the competition ability dependence on the uniqueness.

# Acknowledgement

The studies were supported by the National Natural Science Foundation of China under Grant Nos 10635040 and 70671089.

# Appendix 1

Now we prove that, on the condition that the $x$ is normalized ($0<x_i<1$ and $\sum_{i=1}^{N} x_i = 1$), an SPL function, $P(x) \propto (x-\alpha)^\gamma$, tends to an exponential function if $\alpha \to 1$.

We can write: $P(x) \propto (x-\alpha)^\gamma = (-\alpha)^\gamma (1-x/\alpha)^\gamma = (-\alpha)^\gamma [(1-x/\alpha)^{\alpha/x}]^{\gamma \cdot x/\alpha}$. Since $0<x_i<1$ and $\sum_{i=1}^{N} x_i = 1$, when $\alpha \to 1$, $x_i/\alpha$ is very small. From $\lim_{y \to 0}(1-y)^{1/y} = e$, one has $P(x) \propto (-\alpha)^\gamma [(1-x/\alpha)^{\alpha/x}]^{\gamma \cdot x/\alpha} = (-\alpha)^\gamma e^{\gamma \cdot x/\alpha}$. This is an exponential function.